

\input epsf
\newbox\leftpage \newdimen\fullhsize \newdimen\hstitle
\newdimen\hsbody
\tolerance=1000\hfuzz=2pt
\def\printertype{ps: }
\def\qms{\def\printertype{qms: }
\ifx\answ\bigans\else\voffset=-.4truein\hoffset=.125truein\fi}
\def\bigans{b }
%
\let\answ\bigans
\ifx\answ\bigans\message{This will come out unreduced.}
\magnification=1200\baselineskip=12pt plus 2pt minus 1pt
\hsbody=\hsize \hstitle=\hsize 
%
\else\message{(This will be reduced.} \let\lr=L
\magnification=1000\baselineskip=16pt plus 2pt minus 1pt
\voffset=-.31truein\vsize=7truein\hoffset=-.59truein
\hstitle=8truein\hsbody=4.75truein\fullhsize=10truein\hsize=\hsbody
\output={\ifnum\pageno=0 
  \shipout\vbox{\special{\printertype landscape}\makeheadline
    \hbox to \fullhsize{\hfill\pagebody\hfill}}\advancepageno
  \else
\almostshipout{\leftline%
{\vbox{\pagebody\makefootline}}}\advancepageno
  \fi}
\def\almostshipout#1{\if L\lr \count1=1
\message{[\the\count0.\the\count1]}
      \global\setbox\leftpage=#1 \global\let\lr=R
  \else \count1=2
    \shipout\vbox{\special{\printertype landscape}
      \hbox to\fullhsize{\box\leftpage\hfil#1}}  \global\let\lr=L\fi}
\fi
%
\catcode`\@=11 
\newcount\yearltd\yearltd=\year\advance\yearltd by -1900

%
%

\def\draftmode{\message{ DRAFTMODE }\def\draftdate{{\rm preliminary
draft:
\number\month/\number\day/\number\yearltd\ \ \hourmin}}%
\headline={\hfil\draftdate}\writelabels\baselineskip=20pt plus 2pt
minus 2pt
 {\count255=\time\divide\count255 by 60
\xdef\hourmin{\number\count255}
  \multiply\count255 by-60\advance\count255 by\time
  \xdef\hourmin{\hourmin:\ifnum\count255<10 0\fi\the\count255}}}
\def\nolabels{\def\wrlabel##1{}\def\eqlabeL##1{}\def\reflabel##1{}}
\def\writelabels{\def\wrlabel##1{\leavevmode\vadjust{\rlap{\smash%
{\line{{\escapechar=`
\hfill\rlap{\sevenrm\hskip.03in\string##1}}}}}}}%
\def\eqlabeL##1{{\escapechar-1\rlap{\sevenrm\hskip.05in\string##1}}}%
\def\reflabel##1{\noexpand\llap%
{\noexpand\sevenrm\string\string\string##1}}}
\nolabels
%
\global\newcount\secno \global\secno=0
\global\newcount\meqno \global\meqno=1
\def\newsec#1{\global\advance\secno by1\message{(\the\secno. #1)}
\global\subsecno=0\xdef\secsym{\the\secno.}\global\meqno=1
\noindent{\bf\the\secno. #1}\writetoca{{\secsym} {#1}}
\par\nobreak\medskip\nobreak}
\xdef\secsym{}
\def\eqnres@t{\xdef\secsym{\the\secno.}%
\global\meqno=1\bigbreak\bigskip}
\def\sequentialequations{\def\eqnres@t{\bigbreak}}\xdef\secsym{}
\global\newcount\subsecno \global\subsecno=0
\def\subsec#1{\global\advance\subsecno
by1\message{(\secsym\the\subsecno. #1)}
\bigbreak\noindent{\it\secsym\the\subsecno.
#1}\writetoca{\string\quad
{\secsym\the\subsecno.} {#1}}\par\nobreak\medskip\nobreak}
\def\appendix#1#2{\global\meqno=1\global%
\subsecno=0\xdef\secsym{\hbox{#1.}}
\bigbreak\bigskip\noindent{\bf Appendix #1. #2}\message{(#1. #2)}
\writetoca{Appendix {#1.} {#2}}\par\nobreak\medskip\nobreak}
%
%
\def\eqnn#1{\xdef #1{(\secsym\the\meqno)}\writedef{#1\leftbracket#1}%
\global\advance\meqno by1\wrlabel#1}
\def\eqna#1{\xdef #1##1{\hbox{$(\secsym\the\meqno##1)$}}
\writedef{#1\numbersign1\leftbracket#1{\numbersign1}}%
\global\advance\meqno by1\wrlabel{#1$\{\}$}}
\def\eqn#1#2{\xdef
#1{(\secsym\the\meqno)}\writedef{#1\leftbracket#1}%
\global\advance\meqno by1$$#2\eqno#1\eqlabeL#1$$}
%
\newskip\footskip\footskip10pt plus 1pt minus 1pt 
\def\f@@t{\baselineskip\footskip\bgroup\aftergroup\@foot\let\next}
\setbox\strutbox=\hbox{\vrule height9.5pt depth4.5pt width0pt}
\global\newcount\ftno \global\ftno=0
\def\foot{\global\advance\ftno by1\footnote{$^{\the\ftno}$}}
%
\newwrite\ftfile
\def\footend{\def\foot{\global\advance\ftno by1\chardef\wfile=\ftfile
$^{\the\ftno}$\ifnum\ftno=1\immediate\openout\ftfile=foots.tmp\fi%
\immediate\write\ftfile{\noexpand\smallskip%
\noexpand\item{f\the\ftno:\ }\pctsign}\findarg}%
\def\footatend{\vfill\eject\immediate\closeout\ftfile{\parindent=20pt
\centerline{\bf Footnotes}\nobreak\bigskip\input foots.tmp }}}
\def\footatend{}
%
%
\global\newcount\refno \global\refno=1
\newwrite\rfile
\def\ref{$^{\the\refno}$\nref}
\def\nref#1{\xdef#1{$^{\the\refno}$}\xdef\rfn{\the\refno}
\writedef{#1\leftbracket#1}%
\ifnum\refno=1\immediate\openout\rfile=refs.tmp\fi%
\global\advance\refno by1\chardef\wfile=\rfile\immediate%
\write\rfile{\noexpand\item{{\rfn}.\
}\reflabel{#1\hskip.31in}\pctsign}\findarg}%
\def\findarg#1#{\begingroup\obeylines\newlinechar=`\^^M\pass@rg}%
{\obeylines\gdef\pass@rg#1{\writ@line\relax #1^^M\hbox{}^^M}%
\gdef\writ@line#1^^M{\expandafter\toks0\expandafter{\striprel@x #1}%
\edef\next{\the\toks0}%
\ifx\next\em@rk\let\next=\endgroup\else\ifx\next\empty%
\else\immediate\write\wfile{\the\toks0}%
\fi\let\next=\writ@line\fi\next\relax}}%
\def\striprel@x#1{} \def\em@rk{\hbox{}}

\def\addref#1{\immediate\write\rfile{\noexpand\item{}#1}} 
%

%
\def\startrefs#1{\immediate\openout\rfile=refs.tmp\refno=#1}
\def\xref{\expandafter\xr@f}\def\xr@f[#1]{#1}
\def\refs#1{[\r@fs #1{\hbox{}}]}
\def\r@fs#1{\edef\next{#1}\ifx\next\em@rk\def\next{}\else
\ifx\next#1\xref #1\else#1\fi\let\next=\r@fs\fi\next}
%

%
\newwrite\ffile\global\newcount\figno \global\figno=1
\def\fig{fig.~\the\figno\nfig}
\def\nfig#1{\xdef#1{fig.~\the\figno}%
\writedef{#1\leftbracket fig.\noexpand~\the\figno}%
\ifnum\figno=1\immediate\openout%
\ffile=figs.tmp\fi\chardef\wfile=\ffile%
\immediate\write\ffile{\noexpand\medskip\noexpand\item{Fig.\
\the\figno. }
\reflabel{#1\hskip.55in}\pctsign}\global\advance\figno by1\findarg}
\def\xfig{\expandafter\xf@g}\def\xf@g fig.\penalty\@M\ {}
\def\figs#1{figs.~\f@gs #1{\hbox{}}}
\def\f@gs#1{\edef\next{#1}\ifx\next\em@rk\def\next{}\else
\ifx\next#1\xfig #1\else#1\fi\let\next=\f@gs\fi\next}
\newwrite\lfile
{\escapechar-1\xdef\pctsign{\string\%}\xdef\leftbracket{\string\{}
\xdef\rightbracket{\string\}}\xdef\numbersign{\string\#}}

\def\writestop{\def\writestoppt%
{\immediate\write\lfile{\string\pageno%
\the\pageno\string\startrefs\leftbracket\the\refno\rightbracket%
\string\def\string\secsym\leftbracket\secsym\rightbracket%
\string\secno\the\secno\string\meqno\the\meqno}%
\immediate\closeout\lfile}}
\def\writestoppt{}\def\writedef#1{}
\def\seclab#1{\xdef
#1{\the\secno}\writedef{#1\leftbracket#1}\wrlabel{#1=#1}}
\def\subseclab#1{\xdef #1{\secsym\the\subsecno}%
\writedef{#1\leftbracket#1}\wrlabel{#1=#1}}
\newwrite\tfile \def\writetoca#1{}
\def\leaderfill{\leaders\hbox to 1em{\hss.\hss}\hfill}
\def\writetoc{\immediate\openout\tfile=toc.tmp
   \def\writetoca##1{{\edef\next{\write\tfile{\noindent ##1
   \string\leaderfill {\noexpand\number\pageno} \par}}\next}}}

%
\ifx\answ\bigans
 
scaled\magstep3
 \font\titlei=cmmi10
scaled\magstep3
\font\titleis=cmmi7 scaled\magstep3 \font\titleiss=cmmi5
scaled\magstep3
\font\titlesy=cmsy10 scaled\magstep3 \font\titlesys=cmsy7
scaled\magstep3
\font\titlesyss=cmsy5 scaled\magstep3 
scaled\magstep3
\else
 
scaled\magstep4
 \font\titlei=cmmi10
scaled\magstep4
\font\titleis=cmmi7 scaled\magstep4 \font\titleiss=cmmi5
scaled\magstep4
\font\titlesy=cmsy10 scaled\magstep4 \font\titlesys=cmsy7
scaled\magstep4
\font\titlesyss=cmsy5 scaled\magstep4 
scaled\magstep4
\font\absrm=cmr10 scaled\magstep1 \font\absrms=cmr7 scaled\magstep1
\font\absrmss=cmr5 scaled\magstep1 \font\absi=cmmi10 scaled\magstep1
\font\absis=cmmi7 scaled\magstep1 \font\absiss=cmmi5 scaled\magstep1
\font\abssy=cmsy10 scaled\magstep1 \font\abssys=cmsy7 scaled\magstep1
\font\abssyss=cmsy5 scaled\magstep1 \font\absbf=cmbx10
scaled\magstep1
\skewchar\absi='177 \skewchar\absis='177 \skewchar\absiss='177
\skewchar\abssy='60 \skewchar\abssys='60 \skewchar\abssyss='60
\fi
\skewchar\titlei='177 \skewchar\titleis='177 \skewchar\titleiss='177
\skewchar\titlesy='60 \skewchar\titlesys='60 \skewchar\titlesyss='60
\ifx\answ\bigans\def\abstractfont{\footfont}\else
\def\abstractfont{\def\rm{\fam0\absrm}
\textfont0=\absrm \scriptfont0=\absrms \scriptscriptfont0=\absrmss
\textfont1=\absi \scriptfont1=\absis \scriptscriptfont1=\absiss
\textfont2=\abssy \scriptfont2=\abssys \scriptscriptfont2=\abssyss
\textfont\itfam=\bigit \def\it{\fam\itfam\bigit}
\textfont\bffam=\absbf \def\bf{\fam\bffam\absbf} \rm} \fi
\newfam\mssfam
\font\footmss=cmss8
\font\tenmss=cmss10
\def\tenpoint{\def\rm{\fam0\tenrm}
\textfont0=\tenrm \scriptfont0=\sevenrm \scriptscriptfont0=\fiverm
\textfont1=\teni  \scriptfont1=\seveni  \scriptscriptfont1=\fivei
\textfont2=\tensy \scriptfont2=\sevensy \scriptscriptfont2=\fivesy
\textfont\itfam=\tenit \def\it{\fam\itfam\tenit}
\textfont\mssfam=\tenmss \def\mss{\fam\mssfam\tenmss}
\textfont\bffam=\tenbf \def\bf{\fam\bffam\tenbf} \rm}
%
%
\def\noblackbox{\overfullrule=0pt}
\hyphenation{anom-aly anom-alies coun-ter-term coun-ter-terms}
\def\inv{^{\raise.15ex\hbox{${\scriptscriptstyle -}$}\kern-.05em 1}}

\def\Dsl{\,\raise.15ex\hbox{/}\mkern-13.5mu D} 
\def\dsl{\raise.15ex\hbox{/}\kern-.57em\partial}
\def\del{\partial}

\font\bigit=cmti10 scaled \magstep1
\def\lspace{\ifx\answ\bigans{}\else\qquad\fi}
\def\lbspace{\ifx\answ\bigans{}\else\hskip-.2in\fi} 
\def\boxeqn#1{\vcenter{\vbox{\hrule\hbox{\vrule\kern3pt\vbox{\kern3pt
	\hbox{${\displaystyle #1}$}\kern3pt}\kern3pt\vrule}\hrule}}}
\def\mbox#1#2{\vcenter{\hrule \hbox{\vrule height#2in
		\kern#1in \vrule} \hrule}}  
%

\def\darr#1{\raise1.5ex\hbox{$\leftrightarrow$}\mkern-16.5mu #1}

\def\roughly#1{\raise.3ex\hbox%
{$#1$\kern-.75em\lower1ex\hbox{$\sim$}}}
\font\tenmss=cmss10
\font\absmss=cmss10 scaled\magstep1
\newfam\mssfam
\font\footrm=cmr8  \font\footrms=cmr5
\font\footrmss=cmr5   \font\footi=cmmi8
\font\footis=cmmi5   \font\footiss=cmmi5
\font\footsy=cmsy8   \font\footsys=cmsy5
\font\footsyss=cmsy5   \font\footbf=cmbx8
\font\footmss=cmss8
\def\footfont{\def\rm{\fam0\footrm}
\textfont0=\footrm \scriptfont0=\footrms
\scriptscriptfont0=\footrmss
\textfont1=\footi \scriptfont1=\footis
\scriptscriptfont1=\footiss
\textfont2=\footsy \scriptfont2=\footsys
\scriptscriptfont2=\footsyss
\textfont\itfam=\footi \def\it{\fam\itfam\footi}
\textfont\mssfam=\footmss \def\mss{\fam\mssfam\footmss}
\textfont\bffam=\footbf \def\bf{\fam\bffam\footbf} \rm}
\catcode`\@=12 
%
\newif\ifdraft

\noblackbox
\catcode`\@=11
\newif\iffrontpage
\def\figin{\epsfcheck\figin}\def\figins{\epsfcheck\figins}
\def\epsfcheck{\ifx\epsfbox\UnDeFiNeD
\message{(NO epsf.tex, FIGURES WILL BE IGNORED)}
\gdef\figin##1{\vskip2in}\gdef\figins##1{\hskip.5in}%
\else\message{(FIGURES WILL BE INCLUDED)}%
\gdef\figin##1{##1}\gdef\figins##1{##1}\fi}
\def\DefWarn#1{}
\def\figinsert{\goodbreak\midinsert}
\def\ifig#1#2#3{\DefWarn#1\xdef#1{fig.~\the\figno}
\writedef{#1\leftbracket fig.\noexpand~\the\figno}%
\figinsert\figin{\centerline{#3}}\medskip%
\centerline{\vbox{\baselineskip12pt
\advance\hsize by -1truein\noindent\tensl%
\centerline{{\bf Fig.~\the\figno}~#2}}
}\bigskip\endinsert\global\advance\figno by1}
\ifx\answ\bigans
\def\titleft{\titsm}
\magnification=1200\baselineskip=12pt plus 2pt minus 1pt
%
\voffset=0.35truein\hoffset=0.250truein
\hsize=6.0truein\vsize=8.5 truein
\hsbody=\hsize\hstitle=\hsize
\else\let\lr=L
\def\titleft{\titla}
\magnification=1000\baselineskip=14pt plus 2pt minus 1pt
%
\vsize=6.5truein
\hstitle=8truein\hsbody=4.75truein
\fullhsize=10truein\hsize=\hsbody
\fi
\parskip=4pt plus 15pt minus 1pt
\font\titsm=cmr10 scaled\magstep2
\font\titla=cmr10 scaled\magstep3
\font\tenmss=cmss10
\font\absmss=cmss10 scaled\magstep1
\newfam\mssfam
\font\footrm=cmr8  \font\footrms=cmr5
\font\footrmss=cmr5   \font\footi=cmmi8
\font\footis=cmmi5   \font\footiss=cmmi5
\font\footsy=cmsy8   \font\footsys=cmsy5
\font\footsyss=cmsy5   \font\footbf=cmbx8
\font\footmss=cmss8
\def\footfont{\def\rm{\fam0\footrm}
\textfont0=\footrm \scriptfont0=\footrms
\scriptscriptfont0=\footrmss
\textfont1=\footi \scriptfont1=\footis
\scriptscriptfont1=\footiss
\textfont2=\footsy \scriptfont2=\footsys
\scriptscriptfont2=\footsyss
\textfont\itfam=\footi \def\it{\fam\itfam\footi}
\textfont\mssfam=\footmss \def\mss{\fam\mssfam\footmss}
\textfont\bffam=\footbf \def\bf{\fam\bffam\footbf} \rm}
\def\tenpoint{\def\rm{\fam0\tenrm}
\textfont0=\tenrm \scriptfont0=\sevenrm
\scriptscriptfont0=\fiverm
\textfont1=\teni  \scriptfont1=\seveni
\scriptscriptfont1=\fivei
\textfont2=\tensy \scriptfont2=\sevensy
\scriptscriptfont2=\fivesy
\textfont\itfam=\tenit \def\it{\fam\itfam\tenit}
\textfont\mssfam=\tenmss \def\mss{\fam\mssfam\tenmss}
\textfont\bffam=\tenbf \def\bf{\fam\bffam\tenbf} \rm}
\ifx\answ\bigans\def\abstractfont{\tenpoint}\else
\def\abstractfont{\def\rm{\fam0\absrm}
\textfont0=\absrm \scriptfont0=\absrms
\scriptscriptfont0=\absrmss
\textfont1=\absi \scriptfont1=\absis
\scriptscriptfont1=\absiss
\textfont2=\abssy \scriptfont2=\abssys
\scriptscriptfont2=\abssyss
\textfont\itfam=\bigit \def\it{\fam\itfam\bigit}
\textfont\mssfam=\absmss \def\mss{\fam\mssfam\absmss}
\textfont\bffam=\absbf \def\bf{\fam\bffam\absbf}\rm}\fi
%
\def\f@@t{\baselineskip10pt\lineskip0pt\lineskiplimit0pt
\bgroup\aftergroup\@foot\let\next}
\setbox\strutbox=\hbox{\vrule height 8.pt depth 3.5pt width\z@}
\def\vfootnote#1{\insert\footins\bgroup
\baselineskip10pt\footfont
\interlinepenalty=\interfootnotelinepenalty
\floatingpenalty=20000
\splittopskip=\ht\strutbox \boxmaxdepth=\dp\strutbox
\leftskip=24pt \rightskip=\z@skip
\parindent=12pt \parfillskip=0pt plus 1fil
\spaceskip=\z@skip \xspaceskip=\z@skip
\Textindent{$#1$}\footstrut\futurelet\next\fo@t}
\def\Textindent#1{\noindent\llap{#1\enspace}\ignorespaces}
\def\footnote#1{\attach{#1}\vfootnote{#1}}%

\def\foot{\attach\footsymbolgen\vfootnote{\footsymbol}}
\let\footsymbol=\star
\newcount\lastf@@t           \lastf@@t=-1
\newcount\footsymbolcount    \footsymbolcount=0
\def\footsymbolgen{\relax\footsym
\global\lastf@@t=\pageno\footsymbol}
\def\footsym{\ifnum\footsymbolcount<0
\global\footsymbolcount=0\fi
{\iffrontpage \else \advance\lastf@@t by 1 \fi
\ifnum\lastf@@t<\pageno \global\footsymbolcount=0
\else \global\advance\footsymbolcount by 1 \fi }
\ifcase\footsymbolcount \fd@f\star\or
\fd@f\dagger\or \fd@f\ast\or
\fd@f\ddagger\or \fd@f\natural\or
\fd@f\diamond\or \fd@f\bullet\or
\fd@f\nabla\else \fd@f\dagger
\global\footsymbolcount=0 \fi }
\def\fd@f#1{\xdef\footsymbol{#1}}
\def\space@ver#1{\let\@sf=\empty \ifmmode #1\else \ifhmode
\edef\@sf{\spacefactor=\the\spacefactor}
\unskip${}#1$\relax\fi\fi}
\def\attach#1{\space@ver{\strut^{\mkern 2mu #1}}\@sf}
%
\newif\ifnref
\def\rrr#1#2{\relax\ifnref\nref#1{#2}\else\ref#1{#2}\fi}
\def\ldf#1#2{\begingroup\obeylines
\gdef#1{\rrr{#1}{#2}}\endgroup\unskip}
\def\nrf#1{\nreftrue{#1}\nreffalse}
\def\doubref#1#2{\refs{{#1},{#2}}}

\nreffalse
\def\refout{\footatend\immediate\closeout\rfile\writestoppt
\baselineskip=12pt\hbox{{\bf
References}\hfil}\bigskip{\frenchspacing%
\parindent=12pt\escapechar=` \input
refs.tmp\vfill\eject}\nonfrenchspacing}
%
\def\eqn#1{\xdef #1{(\secsym\the\meqno)}
\writedef{#1\leftbracket#1}%
\global\advance\meqno by1\eqno#1\eqlabeL#1}
\def\eqnalign#1{\xdef #1{(\secsym\the\meqno)}
\writedef{#1\leftbracket#1}%
\global\advance\meqno by1#1\eqlabeL{#1}}
%
\def\chap#1{\newsec{#1}}
\def\chapter#1{\chap{#1}}
\def\sect#1{\subsec{{ #1}}}
\def\section#1{\sect{#1}}
\def\\{\ifnum\lastpenalty=-10000\relax
\else\hfil\penalty-10000\fi\ignorespaces}
\def\note#1{\leavevmode%
\edef\@@marginsf{\spacefactor=\the\spacefactor\relax}%
\ifdraft\strut\vadjust{%
\hbox to0pt{\hskip\hsize%
\ifx\answ\bigans\hskip.1in\else\hskip .1in\fi%
\vbox to0pt{\vskip-\dp
\strutbox\sevenbf\baselineskip=8pt plus 1pt minus 1pt%
\ifx\answ\bigans\hsize=.7in\else\hsize=.35in\fi%
\tolerance=5000 \hbadness=5000%
\leftskip=0pt \rightskip=0pt \everypar={}%
\raggedright\parskip=0pt \parindent=0pt%
\vskip-\ht\strutbox\noindent\strut#1\par%
\vss}\hss}}\fi\@@marginsf\kern-.01cm}
\def\titlepage{%
\frontpagetrue\nopagenumbers\abstractfont%
\hsize=\hstitle\rightline{\vbox{\baselineskip=10pt%
{\abstractfont\pubnum}}}\pageno=0}
\frontpagefalse
\def\pubnum{}
\def\pdate{\number\month/\number\yearltd}
\def\makefootline{\iffrontpage\vskip .27truein
\line{\the\footline}
\vskip -.1truein\leftline{\vbox{\baselineskip=10pt%
{\abstractfont\pdate}}}
\else\vskip.5cm\line{\hss \tenrm $-$ \folio\ $-$ \hss}\fi}
\def\title#1{\vskip .7truecm\titlestyle{\titleft #1}}
\def\titlestyle#1{\par\begingroup \interlinepenalty=9999
\leftskip=0.02\hsize plus 0.23\hsize minus 0.02\hsize
\rightskip=\leftskip \parfillskip=0pt
\hyphenpenalty=9000 \exhyphenpenalty=9000
\tolerance=9999 \pretolerance=9000
\spaceskip=0.333em \xspaceskip=0.5em
\noindent #1\par\endgroup }
\def\autskip{\ifx\answ\bigans\vskip.5truecm\else\vskip.1cm\fi}
\def\author#1{\vskip .7in \centerline{#1}}

\def\address#1{\ifx\answ\bigans\vskip.2truecm
\else\vskip.1cm\fi{\it \centerline{#1}}}
\def\abstract#1{
\vskip .5in\vfil\centerline
{\bf Abstract}\penalty1000
{{\smallskip\ifx\answ\bigans\leftskip 2pc \rightskip 2pc
\else\leftskip 5pc \rightskip 5pc\fi
\noindent\abstractfont \baselineskip=12pt
{#1} \smallskip}}
\penalty-1000}
%

%


\def\bfone{\relax{\rm 1\kern-.35em 1}}
\def\inbar{\vrule height1.5ex width.4pt depth0pt}
\def\IC{{\relax\,\hbox{$\inbar\kern-.3em{\mss C}$}}}
\def\ID{\relax{\rm I\kern-.18em D}}
\def\IF{\relax{\rm I\kern-.18em F}}
\def\IH{\relax{\rm I\kern-.18em H}}
\def\II{\relax{\rm I\kern-.17em I}}
\def\IN{\relax{\rm I\kern-.18em N}}
\def\IP{\relax{\rm I\kern-.18em P}}
\def\IQ{\relax\,\hbox{$\inbar\kern-.3em{\rm Q}$}}
\def\IR{\relax{\rm I\kern-.18em R}}
\font\cmss=cmss10 \font\cmsss=cmss10 at 7pt
\def\ZZ{\relax\ifmmode\mathchoice
{\hbox{\cmss Z\kern-.4em Z}}{\hbox{\cmss Z\kern-.4em Z}}
{\lower.9pt\hbox{\cmsss Z\kern-.4em Z}}
{\lower1.2pt\hbox{\cmsss Z\kern-.4em Z}}\else{\cmss Z\kern-.4em
Z}\fi}
\def\ZZ{\relax\ifmmode\mathchoice
{\hbox{\mss Z\kern-.4em Z}}{\hbox{\mss Z\kern-.4em Z}}
{\lower.9pt\hbox{\mss Z\kern-.4em Z}}
{\lower1.2pt\hbox{\mss Z\kern-.4em Z}}\else{\mss Z\kern-.4em Z}\fi}
\def\CP{\relax\ifmmode\mathchoice
{\hbox{\cmss CP}}{\hbox{\cmss CP}}
{\lower.9pt\hbox{\cmsss CP}}
{\lower1.2pt\hbox{\cmsss CP}}\else{\cmss CP}\fi}
\def\a{\alpha} \def\b{\beta} \def\d{\delta}
 
 \def\l{\lambda}
\def\L{\Lambda}

\def\cF{{\cal F}}

 \def\cM{{\cal M}}

\def\nup#1({{\it Nucl.\ Phys.}\ $\us {B#1}$\ (}
\def\plt#1({{\it Phys.\ Lett.}\ $\us  {#1}$\ (}
\def\cmp#1({{\it Comm.\ Math.\ Phys.}\ $\us  {#1}$\ (}
\def\prp#1({{\it Phys.\ Rep.}\ $\us  {#1}$\ (}
\def\prl#1({{\it Phys.\ Rev.\ Lett.}\ $\us  {#1}$\ (}
\def\prv#1({{\it Phys.\ Rev.}\ $\us  {#1}$\ (}
\def\mpl#1({{\it Mod.\ Phys.\ Let.}\ $\us  {A#1}$\ (}
\def\ijmp#1({{\it Int.\ J.\ Mod.\ Phys.}\ $\us{A#1}$\ (}
\def\tit#1|{{\it #1},\ }
%

%

\def\ni{\noindent}
\def\tilde{\widetilde}

\def\us#1{\bf{#1}}

\def\hyp{{\vrule height 1.9pt width 3.5pt depth -1.5pt}\hskip2.0pt}

\def\Coeff#1#2{{#1\over #2}}
\def\Coe#1.#2.{{#1\over #2}}

\def\coe#1.#2.{\relax{\textstyle {#1 \over #2}}\displaystyle}

\def\shalf{\relax{\textstyle {1 \over 2}}\displaystyle}

\def\to{\rightarrow}
\def\notin{\hbox{{$\in$}\kern-.51em\hbox{/}}}

\def\del{\partial}

\def\nex#1{$N\!=\!#1$}

\def\cc{$^,$}
\def\doubref#1#2{{#1}\cc{#2}}

\catcode`\@=12
%
\newif\ifhypertex
\hypertexfalse
    \def\hyperref#1#2#3#4{#4}
    \def\hyperdef#1#2#3#4{#4}
    \def\hypernoname{}
    \def\e@tf@ur#1{}
    \def\hth/#1#2#3#4#5#6#7{{\tt hep-th/#1#2#3#4#5#6#7}}
\newif\iffigureexists
\newif\ifepsfloaded
\def\epsfcheck{
\ifdraft
\input epsf\epsfloadedtrue
\else
  \openin 1 epsf
  \ifeof 1 \epsfloadedfalse \else \epsfloadedtrue \fi
  \closein 1
  \ifepsfloaded
    \input epsf
  \else
\immediate\write20{NO EPSF FILE --- FIGURES WILL BE IGNORED}
  \fi
\fi
\def\epsfcheck{}}
\def\checkex#1{
\ifdraft
\figureexiststrue
\else\relax
    \ifepsfloaded \openin 1 #1
        \ifeof 1
           \figureexistsfalse
  \immediate\write20{FIGURE FILE #1 NOT FOUND}
        \else \figureexiststrue
        \fi \closein 1
    \else \figureexistsfalse
    \fi
\fi}
\def\missbox#1#2{$\vcenter{\hrule
\hbox{\vrule height#1\kern1.truein
\raise.5truein\hbox{#2} \kern1.truein \vrule} \hrule}$}
\def\lfig#1{
\let\labelflag=#1%
\def\numb@rone{#1}%
\ifx\labelflag\UnDeFiNeD%
{\xdef#1{\the\figno}%
\writedef{#1\leftbracket{\the\figno}}%
\global\advance\figno by1%
}\fi{\hyperref{}{figure}{{\numb@rone}}{Fig.{\numb@rone}}}}
\def\figinsert#1#2#3#4{
\epsfcheck\checkex{#4}%
\def\figsize{#3}%
\let\flag=#1\ifx\flag\UnDeFiNeD
{\xdef#1{\the\figno}%
\writedef{#1\leftbracket{\the\figno}}%
\global\advance\figno by1%
}\fi
\goodbreak\midinsert%
\iffigureexists
\centerline{\epsfysize\figsize\epsfbox{#4}}%
\else%
\vskip.05truein
  \ifepsfloaded
  \ifdraft
  \centerline{\missbox\figsize{Draftmode: #4 not included}}%
  \else
  \centerline{\missbox\figsize{#4 not found}}
  \fi
  \else
  \centerline{\missbox\figsize{epsf.tex not found}}
  \fi
\vskip.05truein
\fi%
{\smallskip%
\leftskip 4pc \rightskip 4pc%
\noindent\footfont \baselineskip=10pt%
{\bf{\hyperdef\hypernoname{figure}{{#1}}{\bf Fig.{#1}}}:~}#2%
\vskip .2truecm
}\endinsert%
}

\def\a{a_1}

\def\b{a_2}

\def\cF{{\cal F}}

\def\g{\gamma}

\def\bifset{\Sigma}

\def\g{\gamma}
\def\CM{{\cal M}_0}

\def\simpA{W_{\!A_{n-1}}}
\def\bifset{\Sigma}
\def\g{\gamma}

\def\cF{{\cal F}}

\def\a{\alpha}
\def\b{\beta}

\def\L{\Lambda}
\def\CM{{\cal M}_0}

\def\ln{{\,\log\,}}
\def\cap{\circ}
%
\ldf\SW{N.\ Seiberg and E.\ Witten, \nup426(1994) 19, \hth/9407087;
\nup431(1994) 484, \hth/9408099.}
\ldf\KLTYa{A.\ Klemm, W.\ Lerche, S.\ Theisen and S.\ Yankielowicz,
\plt B344(1995) 169, \hth/9411048.}
\ldf\KLT{A.\ Klemm, W.\ Lerche and S.\ Theisen, {\it Nonperturbative
Effective Actions of N=2 Supersymmetric Gauge Theories}, preprint
CERN-TH/95/104, LMU-TPW 95-7, \hth/9505150.}
\ldf\AF{P. Argyres and A. Faraggi, \prl 73 (1995) 3931,
\hth/9411057.}
\ldf\Arn{See e.g., V.\ Arnold, A.\ Gusein-Zade and A.\ Varchenko,
{\it Singularities of Differentiable Maps I, II}, Birkh\"auser 1985.}
\ldf\thooft{G.\ `t Hooft, \nup190(1981) 455.}
\ldf\HT{C.\ Hull and
P.\ Townsend, {\it Unity of Superstring Dualities}, preprint
QMW-94-30, \hth/9410167.}
\ldf\MDSS{M.\ Douglas and S.\ Shenker, {\it
Dynamics of SU(N) Supersymmetric Gauge Theory}, preprint RU-95-12,
\hth/9503163.}
\ldf\Stro{A.\ Strominger, {\it Massless Black Holes and Conifolds in
String Theory}, ITP St.\ Barbara preprint, \hth/9504090.}
\ldf\WiVa{E.\ Witten, {\it String Theory Dynamics In Various
Dimensions}, Princeton preprint, \hth/9503124; C.\ Vafa, unpublished,
as cited therein.}
\ldf\DS{U.\ Danielsson and B.\ Sundborg, {\it The Moduli Space and
Monodromies of N=2 Supersymmetric SO(2r+1) Yang-Mills Theory},
preprint USITP-95-06, UUITP-4/95, \hth/9504102.}
\ldf\PAMD{P.\ Argyres and M.\ Douglas, {\it
New Phenomena in SU(3) Supersymmetric Gauge Theory}, preprint
IASSNS-HEP-95/31, RU-95-28, \hth/9505062.}
\ldf\Nf{A.\ Hanany and Y.\ Oz, {\it On the Quantum Moduli Space of
N=2 Supersymmetric SU(Nc) Gauge Theories},
preprint TAUP-2248-95,WIS-95/19/May-PH, \hth/9505075;
P.\ Argyres, M.\ Plesser and A. Shapere, {\it The Coulomb Phase of
N=2 Supersymmetric QCD}, preprint IASSNS-HEP-95/32, UK-HEP/95-06,
\hth/9505100.}
\ldf\KaVa{S.\ Kachru and C.\ Vafa, {\it Exact Results for N=2
Compactifications of Heterotic Strings}, preprint HUTP-95/A016,
\hth/9505105.}
\ldf\LW{W.\ Lerche and N.P.\ Warner, \nup358 (1991) 571.}
\ldf\DVV{R.\ Dijkgraaf, E. Verlinde and H. Verlinde, \nup{352} (1991)
59.}
\ldf\BL{A.\ Brandhuber and K.\ Landsteiner, {\it On the Monodromies
of N=1 Supersymmetric Yang-Mills Theory with Gauge Group SO(2n)},
preprint CERN-TH/95-180, \hth/9507008.}
\ldf\MO{C.\ Montonen and D.\ Olive, \plt72 (1977) 117.}
\ldf\MW{D.\ Olive and E.\ Witten, \plt78 (1978) 97.}
\ldf\FHSV{S. Ferrara, J. A. Harvey, A. Strominger and C. Vafa, {\it
Second-Quantized Mirror Symmetry}, preprint EFI-95-26, \hth/9505162.}
\ldf\AM{P.\ Aspinwall and D.\ Morrison, {\it U-Duality and Integral
Structures}, preprint CLNS-95-1334, \hth/9505025.}
\ldf\GMS{B.\ Greene, D.\ Morrison and A.\ Strominger, {\it Black Hole
Condensation and the Unification of String Vacua}, preprint
CLNS-95-1335, \hth/9504145.}
%
%
\voffset=0.00truein\hoffset=0.150truein
\hsize=6.0truein\vsize=8.6 truein
%
%
\def\abstr{We review some simple group theoretical properties of BPS
states, in relation with the singular homology of level surfaces.
Primary focus is on classical and quantum \nex2 supersymmetric
Yang-Mills theory, though the considerations can be applied
to string theory as well.}
\font\eightrm=cmr8

\font\titsm=cmr10 scaled\magstep2
\nopagenumbers
\def\pubnum{
\hbox{CERN-TH/95-183}
\hbox{hepth@xxx/9507011}}
\def\pdate{
\hbox{CERN-TH/95-183}
\hbox{July 1995}
}
\titlepage
\vskip 2.5truecm
\title{\titsm BPS States, Weight Spaces and Vanishing Cycles}
\bigskip
\bigskip
\bigskip
\tenpoint
\font\eightrm=cmr8

\centerline{W.\ Lerche}
\bigskip
\centerline{{\it CERN, CH 1211 Geneva 23, Switzerland}}
\bigskip
\vfil
{\centerline{\it Contribution to the Proceedings of }}
{\centerline{\it Strings '95, USC, Los Angeles,}}
{\centerline{\it and to the Proceedings of the Trieste Conference on
}}
{\centerline{\it S-Duality and Mirror Symmetry, 1995}}
\bigskip
\bigskip\vfil
\noindent{\tenrm \abstr}
\vskip 3.truecm
\eject

\def\pdate{}
\hsize=6truein \vsize=8.6truein
\voffset=.375truein  \hoffset=.25truein
\font\eightrm=cmr8

\font\eightit=cmti8

\nopagenumbers
\centerline{{ \tenbf BPS STATES, WEIGHT SPACES AND
VANISHING CYCLES}}
\vskip.8truecm
\centerline{{\eightrm W.\ LERCHE}}
\centerline{{\eightit CERN, Geneva, Switzerland}}

\vskip1.3truecm
\vbox{\hbox{\centerline{{\eightrm ABSTRACT}}}
{\smallskip\leftskip 1.truecm \rightskip 1.truecm \noindent \footfont
\baselineskip=10pt \abstr\smallskip}}

\footline={\hss\tenrm\folio\hss}

\baselineskip=12pt plus 2pt minus 1pt
\sequentialequations

 There has been dramatic recent progress in understanding
non-perturbative properties of certain supersymmetric field
$\!\!$\nrf{\SW\KLTYa\AF\MDSS\PAMD\DS\KLT\Nf\BL}{\SW}$^{\!-\!}${\BL}\
and string
$\!\!$\nrf{\HT\WiVa\Stro\GMS\AM\KaVa\FHSV}{\HT}$^{\!-\!}${\FHSV}\
theories. A common feature of such theories are singularities in the
quantum moduli space, arising from certain BPS states becoming
massless in these vacuum parameter regions. Much of the present
discussion is centered at the properties of such states. Such states
typically have various different representations, for example, a
representation in terms of ordinary elementary fields, or, in a dual
formulation, in terms of non-perturbative solitonic bound states\MO.

We will first discuss some generic properties of BPS states, with
main emphasis on \nex2 supersymmetric theories. We will then
concretely specialize further below to classical and quantum
Yang-Mills theory.

The mass of a BPS state is directly given\MW\ in terms of the central
charge $Z$ of the relevant underlying \nex2 or \nex4 supersymmetry
algebra,
$$
m^2\ \simeq\ |Z|^2\ ,\qquad\ \
Z\ \equiv\ \vec q\cdot\vec a + \vec g\cdot\vec a_D\ .
\eqn\Zaad
$$
Here, $\vec q,\vec g$ are the electric and magnetic charges of the
state in question, and $\vec a,\vec a_D$ are the classical values of
the Higgs field and ``magnetic dual'' Higgs field, respectively. An
important insight\SW\ is that $Z$ can be represented as a period
integral of some prime form $\l$ on a suitable ``level'' surface $X$,
ie.,
$$
Z\ =\ \int_\nu\l\ ,
\eqn\Zl
$$
where $\nu$ is a cycle in the middle homology of $X$, $\nu\in
H_{dim_C(X)}(X,\ZZ)$. Obviously, $Z=0$ if $\nu=0$ (since $\l$ does
not blow up), so that the masslessness of a BPS state can be
attributed to a collapsed ``vanishing cycle'' on $X$. Such vanishing
cycles shrink to zero in certain regions $\Sigma(X)$ of the moduli
space $\cM(X)$, and it was the idea of Seiberg and Witten\SW\ to
literally identify $\cM(X)$ with the quantum moduli space of the
physical model in question. Typically, the singular locus $\Sigma(X)$
consists of branches of codimension one, and on any such branch just
one cycle degenerates, see \lfig\figdegen; of course, on
intersections of such branches the surface $X$ becomes singular in a
more non-trivial way.

\figinsert\figdegen{On the singular locus $\Sigma(X)$ in moduli space
$\cM$, the level surface $X$ degenerates by pinching of vanishing
cycles $\nu$. The coordinates of any such cycle with respect to some
symplectic basis of $H_{dim_C(X)}(X,\ZZ)$ gives the electric and
magnetic quantum numbers of the corresponding massless BPS state.
Shown here is the genus two Riemann surface associated with $SU(3)$
\nex2 supersymmetric Yang-Mills theory.}{1.8in}{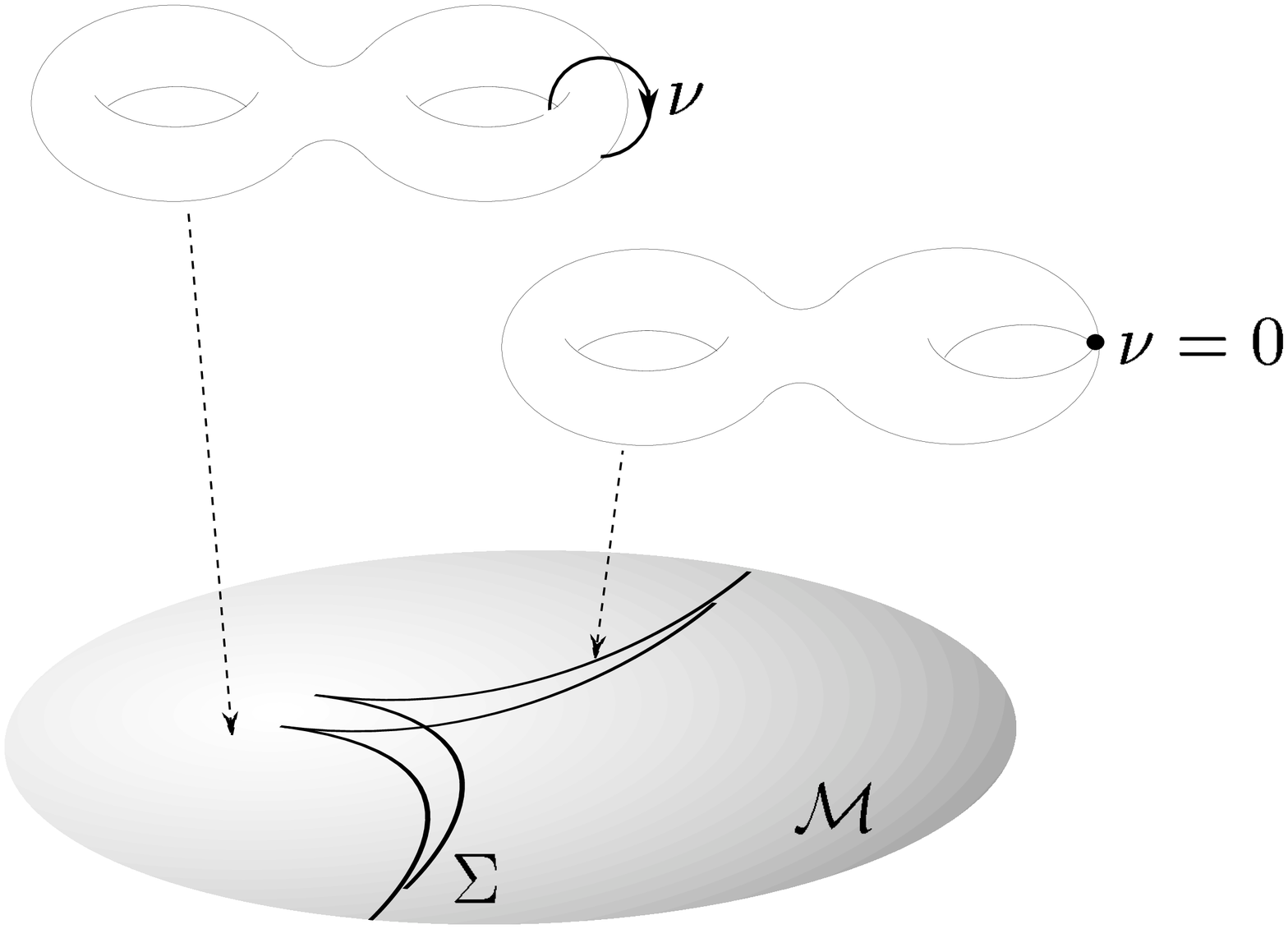}

It turns out that many features of the massless BPS spectrum can
directly be studied in terms of the singular homology of $X$. Note,
however, that various properties crucially depend on whether $d\equiv
dim_C(X)$ is even or odd.

Odd $d$ is the situation of quantum \nex2 supersymmetric Yang-Mills
theory, where $X$ is a hyperelliptic curve\doubref\KLTYa\AF, and of
\nex2 supersymmetric type II string compactifications, where $X$ is a
Calabi-Yau manifold\doubref\Stro\GMS$^{\!,\!}${\KaVa}. In these
theories, the massless BPS states are given by ``matter''
hypermultiplets. For this case of odd $d$, the intersection form
$\Omega$ of $H_d$ is skew-symmetric, so that there is a natural split
of $H_d$ into two sets of non-intersecting cycles, whose bases may be
denoted by $\g_\a$ and $\g_\b$. This split just corresponds to
distinguishing electric and magnetic degrees of freedom.
Specifically, a vanishing cycle can be expanded as follows:
$$
\nu\ =\ \vec q\cdot\vec \g_\a + \vec g\cdot\vec \g_\b\,,
\qquad\qquad \g_\a,\g_\b\in H_{dim_C(X)}(X,\ZZ)\ .
\eqn\nuexp
$$
Identifying $\vec a=\int_{\vec\g_\a}\!\!\l$, $\vec
a_D=\int_{\vec\g_\b}\!\!\l$, we immediately see that the electric and
magnetic quantum numbers $\vec q,\vec g$ of a massless BPS state are
simply given by the coordinates of the corresponding vanishing
cycle\KLT. Obviously, under a change of homology basis, the charges
change as well, but this is nothing but a duality rotation. What
remains invariant is the intersection number
$$
\nu_i\cap\nu_j\ =\ \vec\nu^t\cdot\Omega\cdot\vec\nu\
= \langle\vec g_i, \vec q_j\rangle-\langle\vec g_j, \vec q_i\rangle\
\in\
\ZZ\ , \eqn\dzw
$$
where $\Omega$ is a symplectic metric, and $\langle\ ,
\ \rangle$ is the inner product in weight space. (If we take both
electric and magnetic charge vectors in simple root bases, then
$\Omega=\big(\!{0\atop-C}{C\atop 0}\!\big)$, where $C$ is a Cartan
matrix, or some generalization of it).

Note that \dzw\ represents the well-known Dirac-Zwanziger
quantization condition for the possible electric and magnetic
charges, and we see that it satisfied by construction. The vanishing
of the r.h.s.\ of \dzw\ is required for two states to be local with
respect to each other\doubref\thooft\PAMD. Thus, only states that
are related to non-intersecting cycles are mutually local, and can be
simultaneously represented in a local effective lagrangian. For
example, 't Hooft-Polyakov monopoles (typically associated with
$\g_\b$-cycles) are not local with respect to elementary gauge bosons
(associated with the dual, intersecting $\g_\a$-cycles).

Quintessential for the solution of \nex2 Yang-Mills theory\SW\ was
the study of the global non-abelian monodromy properties of the
quantum moduli space. Specifically, we noted that there is a
particular vanishing cycle $\nu$ associated with each branch of
$\Sigma(X)$. The monodromy action on any given cycle, $\g\in
H_d(X,\ZZ)$, is directly determined in terms of this vanishing cycle
by means of the Picard-Lefshetz formula\Arn:
$$
M_\nu:\ \ \ \d\ \longrightarrow\ \d - (\d\cap\nu)\,\nu\ ,
\eqn\PicLef
$$
where $\cap$ is the intersection product. From this one can find for
a cycle of the form \nuexp\ the following monodromy matrix\KLT:
$$
{M}_{\nu=(g, q)}= \pmatrix{\bfone + \vec q\otimes \vec g& \vec
q\otimes \vec q\cr -\vec g\otimes \vec g&\bfone -\vec g\otimes \vec
q}\ . \eqn\monmatrix
$$
It directly expresses the correct logarithmic shift property of the
\nex2 quantum effective action $\cF(a)$, and thus automatically
guarantees a consistent physical picture. That is, near the vanishing
of some $\nu$, the monodromy shift of the gauge coupling
$\tau\equiv\del^2_a \cF(a)$, when expressed in suitable local
variables, is
$$
\Delta_\nu \tau_{ij}\ =
{\del\over\del a_i}\Delta_\nu a_{D,j} \ =\
-(\g_j^*\cap\nu){\del\over\del a_i}Z_\nu\ =\   -\nu_i\,\nu_j\ ,
\eqn\tauhshft
$$
where $\g^*$ is the cycle dual to $\g$. This is indeed precisely the
monodromy associated with the corresponding one-loop \nex2 effective
action in the local patch near the singular line $\Sigma(X)^{(\nu)}$:
$$
\cF_\nu\ =\ \Coeff1{4\pi i}{Z_\nu}^2\ln\big[{Z_\nu\over\L}\big]\ .
$$

On the other hand, if $d$ is even, the intersection metric $\Omega$
is symmetric, and there is no natural distinction between
``electric'' and ``magnetic'' cycles (typically, $\Omega$ is given by
a Cartan matrix). This is essentially the classical, self-dual
situation, where there are non-abelian gauge bosons among the
massless BPS states. Specifically, $d=0$ corresponds to classical
\nex2 Yang-Mills theory, and $d=2$ corresponds to type II
supersymmetric string compactifications\doubref\HT\WiVa$^,$\AM,
with $X=K_3$.

Most of the above formulas apply just as well (up to a sign change in
the PL formula), if we simply drop the magnetic variables $a_D,\g_b$.
Since now the monodromy matrices do not link electric and magnetic
sectors, the PL monodromy does not describe block off-diagonal
logarithmic shifts that were attributed before to perturbative
corrections, but rather gives directly discrete gauge (Weyl group)
transformations -- this indeed reflects properties of a theory
without quantum corrections.

To summarize the general scheme, we present the following
suggestive table, where $C$ denotes the Cartan matrix, $\L_{R,W}$
root and weight lattices, $W$ a simple singularity, and a canonical
basis of the vanishing cycles has been chosen. One clearly recognizes
the alternating pattern depending on whether $d$ is even or odd.

$$
\vbox{\offinterlineskip\tabskip=0pt
\halign{\strut\vrule#
&\hfil$#$
&\vrule#
&{}~~$#$~~\hfil
&$#$~~\hfil
&~$#$~~\hfil
&$#$~~\hfil
&~~$#$\hfil
&\vrule#
\cr
\noalign{\hrule}
& {\rm phys.\ model} &
&X
& d(X)
& H_d(X,\ZZ)
& {\rm  int.\ form}
& {\rm  BPS\ states}
&\cr
\noalign{\hrule}
& {\rm class.\ SYM} &
& W\!=\!0
& 0
& \L_R
& C
& {\rm gauge\ bosons}
&\cr
& {\rm quant.\ SYM} &
& {W}^2\!-\!1\!=\!y^2
& 1
& \L_W\otimes\L_R
& \big(\!{0\atop-C}{C\atop 0}\!\big)
& {\rm dyons}
&\cr
& \,{\rm type\ II\ comp} &
& K_3
& 2
& {\L_R(E_8)}^2\!\!\otimes\! {\L_{1,1}}^3
& {C_{E_8}}^2\!\otimes\!{C_{1,1}}^3
& {\rm gauge\ solitons}
&\cr
& \,{\rm type\ II\ comp} &
& {\rm Calabi\hyp Yau }
& 3
& ...
& ...
& {\rm dyonic\ bl\ holes}
&\cr
\noalign{\hrule}}
\hrule}
$$

Since classical \nex2 Yang-Mills theory is the simplest
example, we like now to discuss it in somewhat more detail\KLT,
focusing mainly on gauge group $G=SU(n)$.

The scalar superfield component $\phi$ labels a continuous family of
inequivalent ground states, which constitutes the classical moduli
space, $\cM_0$. One can always rotate it into the Cartan sub-algebra,
$
\phi=\sum_{k=1}^{n-1}a_k H_k
$,
with $H_k=E_{k,k}-E_{k+1,k+1},\,
(E_{k,l})_{i,j}=\delta_{ik}\delta_{jl}$. For generic eigenvalues of
$\phi$, the $SU(n)$ gauge symmetry is broken to the maximal torus
$U(1)^{n-1}$, whereas if some eigenvalues coincide, some larger,
non-abelian group $H\subseteq G$ remains unbroken. Precisely which
gauge bosons are massless for a given background $\vec a=\{a_k\}$,
can easily be read off from the central charge formula:
$Z_q(a)\equiv \vec q\cdot \vec a$, where we take for the charge
vectors
$\vec q$ the roots $\vec\alpha\in\L_R(G)$ in Dynkin basis.

The Cartan sub-algebra variables $a_k$ are not gauge invariant and in
particular not invariant under discrete Weyl transformations.
Therefore, one introduces other variables for parametrizing the
classical moduli space, which are given by the Weyl invariant
Casimirs
$u_k(a)$. These variables parametrize the Cartan sub-algebra modulo
the Weyl group, ie, $\{u_k\}\cong \IC^{n-1}/S(n)$, and can be
obtained by a Miura transformation:
$$
\prod_{i=1}^n\big(x-Z_{\l_i}(a)\big)\ =\
x^n-\sum_{l=0}^{n-2}u_{l+2}(a)\,x^{n-2-l}
\ \equiv\ \simpA(x,u)\ .
\eqn\WAnSing
$$
Here, $\l_i$ are the weights of the $n$-dimensional fundamental
representation, and $\simpA(x,u)$ is nothing but the simple
singularity\Arn\ associated with $SU(n)$, with
$$
u_k(a)\ =\ (-1)^{k+1}\sum_{j_1\not=...
\not=j_k}Z_{\l_{j_1}}Z_{\l_{j_2}}\dots Z_{\l_{j_k}}(a)\ .\eqn\sympol
$$
These symmetric polynomials are manifestly invariant under the Weyl
group $S(n)$, which acts by permutation of the weights $\l_i$.

{}From the above we know that whenever $Z_{\l_i}(a)=Z_{\l_j}(a)$ for
some $i$ and $j$, there are, classically, extra massless non-abelian
gauge bosons, since $Z_{ \a}=0$ for some root $\a$. For such
backgrounds the effective action becomes singular. The classical
moduli space is thus given by the space of Weyl invariant
deformations modulo such singular regions:
$\CM=\{u_k\}\backslash\bifset_0$. Here, $\bifset_0\equiv\{u_k:
\Delta_0(u_k)=0\}$ is the zero locus of the ``classical''
discriminant
$$
\Delta_0(u)\ =\
\prod_{i<j}^n(Z_{\l_i}(u)-Z_{\l_j}(u))^2\ =\
\prod_{{{\rm positive}\atop{\rm roots}\
\alpha}}\!\!(Z_{\alpha})^2(u)\ ,
\eqn\cdiscdefu
$$
of the simple singularity \WAnSing. We schematically depicted the
singular loci $\bifset_0$ for $n=2,3,4$ in \lfig\figdisc.

\figinsert\figdisc{Singular loci $\bifset_0$ in the classical moduli
spaces $\CM$ of pure $SU(n)$ \nex2 Yang-Mills theory. They are
nothing but the bifurcation sets of the type $A_{n-1}$ simple
singularities, and reflect all possible symmetry breaking patterns in
a gauge invariant way (for $SU(3)$ and $SU(4)$ we show only the real
parts). The picture for $SU(4)$ is known in singularity theory as the
``swallowtail''. }{1.0in}{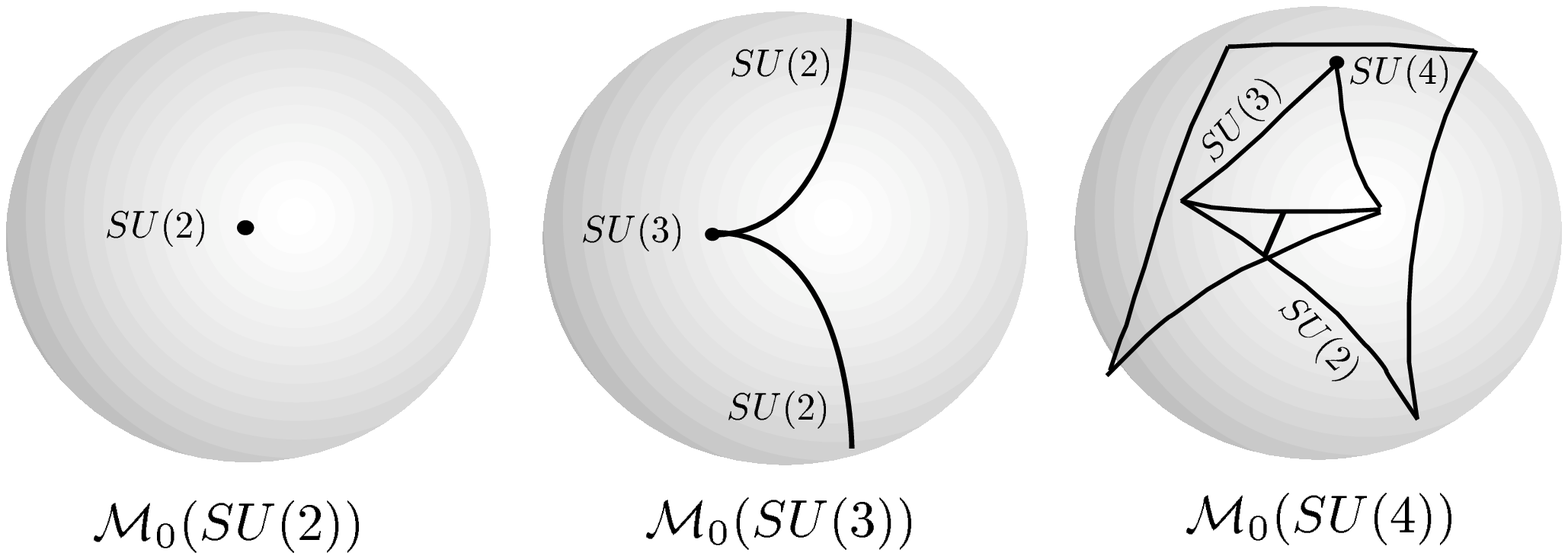}

The discriminant loci $\bifset_0$ are generally given by intersecting
hypersurfaces of complex codimension one. On each such surface one
has $Z_{\alpha}=0$ for some pair of roots $\pm\a$, so that there is
an unbroken $SU(2)$. Furthermore, since $Z_{\alpha}=0$ is a fixed
point of the Weyl transformation $r_{\a}$, the Weyl group action is
singular on these surfaces. On the intersections of these surfaces
one has, correspondingly, larger unbroken gauge groups. All planes
together intersect in just one point, namely in the origin, where the
gauge group $SU(n)$ is fully restored. Thus, what we learn is that
all possible classical symmetry breaking patterns are encoded in the
discriminants of $\simpA(x,u)$.

For classical $SU(n)$ \nex2 Yang-Mills theory, the relevant level
surface $X$ is zero dimensional and given by the following set of
points:
$$
X\ =\ \big\{\,x\,:\,\simpA(x,u)=0\,\big\}\ =\
\big\{\,Z_{\l_i}(u)\,\big\}\ .
\eqn\levsurf
$$
It is singular if any two of the $Z_{\l_i}(u)$ coincide, and indeed,
the vanishing cycles are just given by the differences: $\nu_\a
=Z_{\l_i}-Z_{\l_j}=Z_{\a}$, ie., by the central charges associated
with the non-abelian gauge bosons. It is indeed well-known\Arn\ that
$\nu_\a$ generate the root lattice: $H_0(X,\ZZ)\cong\L_R$. We
depicted the level surface for $G=SU(3)$ in \lfig\figclassYM.

\figinsert\figclassYM{Level manifold for classical $SU(3)$ Yang-Mills
theory, given by points in the $x$-plane;  they form the
projection of a weight diagram. The dashed lines are the vanishing
cycles associated with non-abelian gauge bosons (having corresponding
quantum numbers, here in Dynkin basis). The masses are proportional
to the
lengths of the lines and thus vanish if the cycles
collapse.}{1.4in}{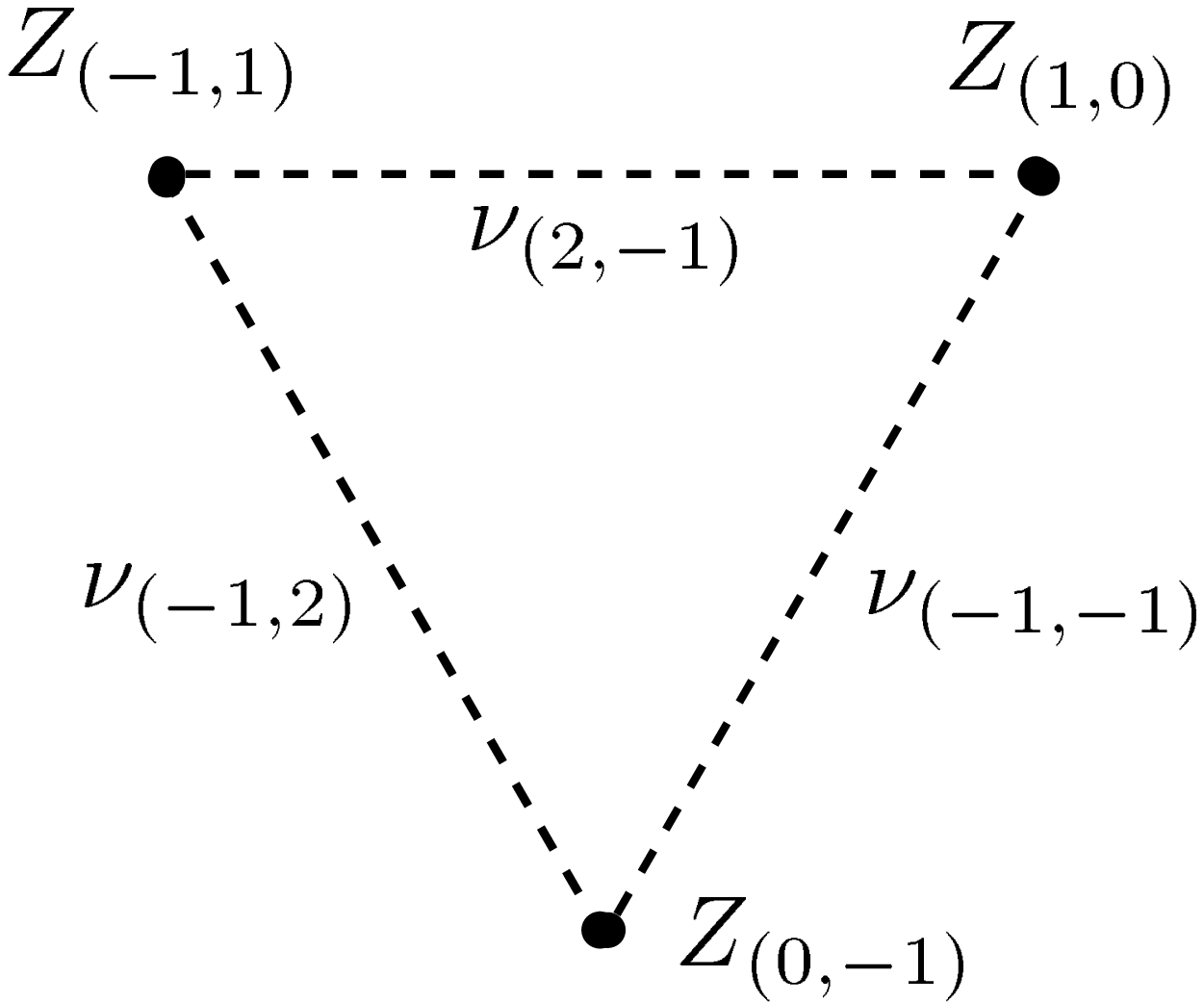}

Pictures like this one have a very concrete group theoretical
meaning. In fact, if we choose (as in \lfig\figclassYM) a special
region in the moduli space where only the top Casimir, $u_n$, is
non-zero, the picture becomes precisely the projection of the weights
$\l_i$ (that live in the $n-1$-dimensional weight space) to the
two-dimensional eigenspace of the Coxeter element with $\ZZ_n$
action (why this so is follows from the considerations of ref.\LW,
where group theoretical aspects of $u_k=0, k\not=n$ are discussed).

Thus, we see another manifestation of the close
connection between the vanishing homology of $X$ and $SU(n)$ weight
space. The intersection numbers
of the vanishing cycles are just given by the inner products between
root vectors, $\nu_{\a_i}\cap\nu_{\a_j} = \langle\a_i,\a_j\rangle$
(self-intersections counting $+2$), and the Picard-Lefshetz formula
\PicLef\ coincides in this case with the well-known formula for Weyl
reflections, with matrix representation:
$
M_{\a_i} = \bfone - \a_i\otimes w_i
$
(where $w_i$ are the fundamental weights).

Note that the corresponding situation in classical string
theory\doubref\HT\WiVa\ is when one takes $X=K_3$ and replaces the
euclidean $SU(n)$ weight space with the lorentzian Narain lattice,
which is isomorphic to the lattice of 2-cycles of the $K_3$ surface.

Most of the above considerations apply more or less directly to the
other simply laced Lie groups of type $D$ and $E$, for which the
following simple singularities \Arn\ are relevant:\foot {For
non-simply laced gauge groups $B_n$, the corresponding boundary
singularities are relevant; cf., see the discussion of the quantum
theory in ref.\DS.}
\goodbreak
$$
\eqalign{
W_{D_n}(x_1,x_2,u)&={x_1}^{n-1}+\shalf x_1\, {x_2}^2
-\sum_{l=1}^{n-1}u_{2l}\,{x_1}^{n-l-1}-\tilde u_n x_2\cr
W_{E_6}(x_1,x_2,u)&={x_1}^3+{x_2}^4 - u_2 x_1 x_2^2-u_5 x_1 x_2-u_6
x_2^2 -u_8 x_1-u_9 x_2-u_{12} \cr
W_{E_7}(x_1,x_2,u)&={x_1}^3+x_1{x_2}^3 - u_2 x_1^2 x_2-u_6 x_1^2-u_8
x_1 x_2 -u_{10} x_2^2\cr&\ \ -u_{12} x_1-u_{14} x_2-u_{18}\cr
W_{E_8}(x_1,x_2,u)&={x_1}^3+{x_2}^5 - u_2 x_1 x_2^3-u_8 x_1 x_2^2-
u_{12} x_2^3-u_{14} x_1 x_2\cr&\ \ -u_{18} x_2^2-u_{24} x_2-u_{30}\ ,
\cr
}\eqn\DEsing
$$
The discriminants of these singularities give indeed precisely the
singularities in the corresponding classical Yang-Mills moduli
spaces.

Note, though, that the number of variables of these
singularities is two, and not one as for $A_{n-1}$. Certain features,
like the intersection properties of vanishing cycles, depend
critically on the number of variables. One usually stabilizes the
situation by adding irrelevant quadratic pieces to the singularities,
so that all ADE singularities are represented by three variables.
This would however confuse the situation in the present context, and
one
rather prefers to have only one variable,~$x$.

For $D_n$ this can easily be achieved\BL\ by integrating out $x_2$
using its ``equations of motion'', $\del_{x_2}W=0$ (just as it is
known from the LG description of type $D_n$ \nex2 superconformal
minimal
models\DVV). The resulting level surface,
$$
X:\ \ \tilde W_{D_n}(x)\equiv
x^2W_{D_n}(x_1=x^2,x_2=\tilde u_n/x^2) \ =\ 0\ ,
$$
gives (for $u_l=0, l\not=2n-2$) indeed the projection of the weight
diagram of the vector representation, see \lfig\DsixPolytope\ for an
example. Note that the correct interpretation of these surfaces is
more involved. Specifically, even though the origin $x=0$ has a
double zero, no massless gauge bosons are associated with it, because
this ``zero distance'' does not correspond to a projection of a root
vector (the two weights that project to the origin differ by sums of
roots, and this rather corresponds to a multi-particle state). The
correct surface is obtained by modding out by $x\to -x$, as each root
appears twice in the picture. Taking this into account by
appropriately considering only $\ZZ_2$-odd linear combinations of
cycles, one obtains the correct singular behavior.\BL

\vskip-.3cm
\figinsert\DsixPolytope{Level surface for classical $D_6$ Yang-Mills
theory, in
analogy to \lfig\figclassYM. It takes this Coxeter-symmetric form if
only the top Casimir is non-zero. The lines represent the vanishing
cycles and are projections of root vectors. Note that each of the
30 positive roots occurs twice, and that the center is two-fold
degenerate. This kind of pictures has appeared before in the
discussion of integrable two-dimensional \nex2 LG models\LW, where,
precisely as here, the line lengths give the masses of BPS states
(they are given by the components of the Frobenius-Perron eigenvector
of the Cartan matrix).}
{1.8in}{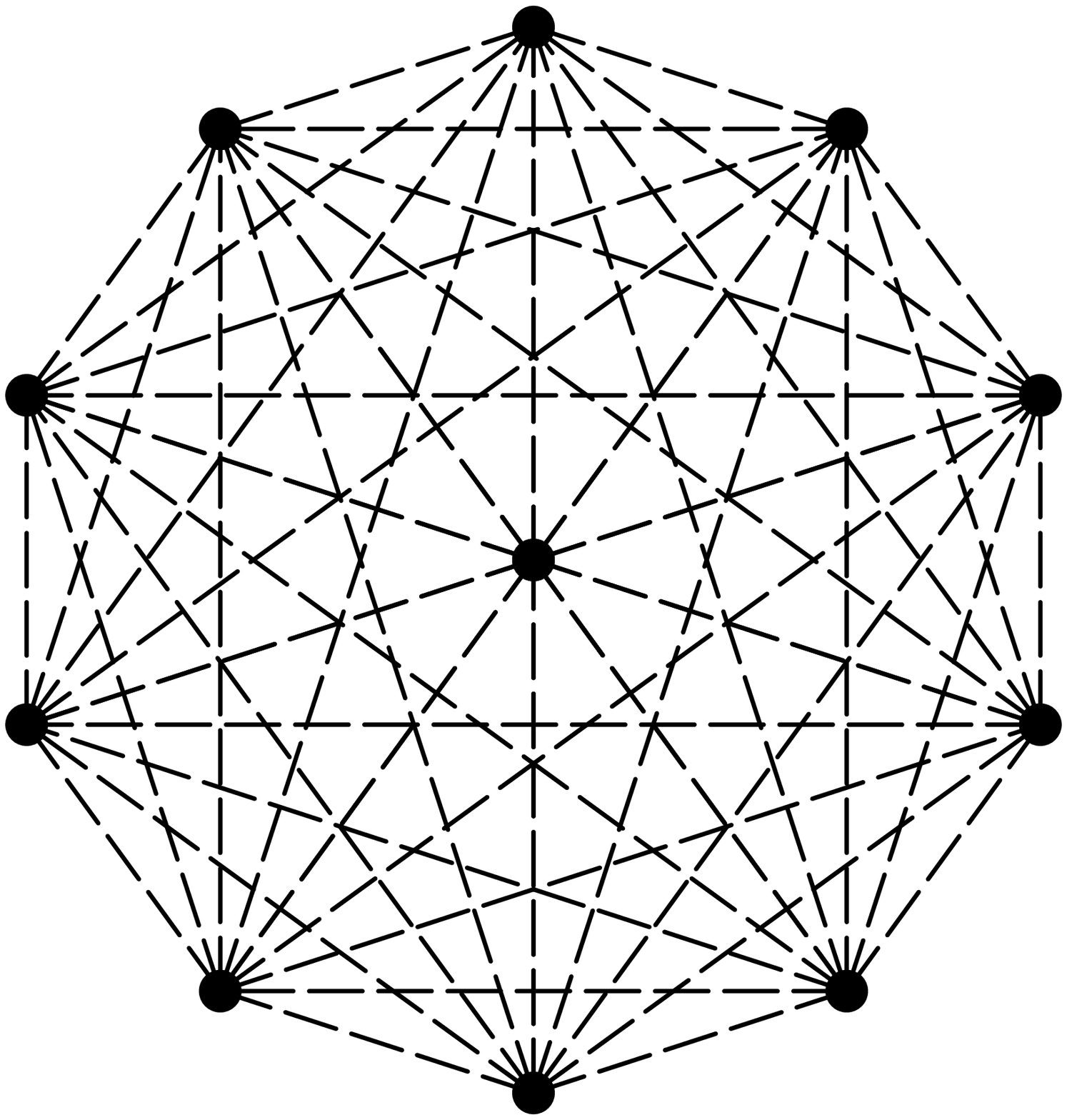}

 How this is precisely to be done for the exceptional groups is
less clear at the moment, but in analogy one expects that after
eliminating $x_2$ from the exceptional simple singularities, one
obtains expressions $\tilde W_{E_n}(x,u)$ whose orders are given by
the dimensions of the defining representation, and which might have a
close relationship to Lax operators obtained from Drinfel'd-Sokolov
reduction. If true, the level surfaces for $E_n$
would look like the pictures given in ref.\LW.

We now turn to the quantum version of the \nex2 Yang-Mills theories,
where the issue is to construct curves $X$ whose moduli spaces
$\cM_\L$ give the supposed quantum moduli spaces. We have seen that
the classical theories are characterized by simple singularities,
so we may expect that the quantum versions should also have something
to do with them. Indeed, for $G=SU(n)$ the appropriate manifolds
were found in\doubref\KLTYa\AF\ and are given by
$$
X:\ \ y^2 \ =\ \left(W_{A_{n-1}}(x,u_i)\right)^2-\Lambda^{2n}\
\ ,\eqn\aaa
$$
which corresponds to special genus $g=n-1$ hyperelliptic curves.
Above, $\L$ is the dynamically generated quantum scale.

Since $y^2$ factors into $W_{A_{n-1}}\pm\L^n$, the situation is in
some respect like two copies of the classical theory, with the top
Casimir $u_n$ shifted by $\pm\L^n$. Specifically,  the
``quantum'' discriminant, whose zero locus $\bifset_\L$ gives the
singularities in the quantum moduli space $\cM_\L$, is easily seen to
factorize as follows:
$$
\eqalign{
\Delta_\Lambda(u_k,\L)\ &\equiv\
\prod_{i<j}(Z_{\l_i}^+-Z_{\l_j}^+)^2(Z_{\l_i}^--Z_{\l_j}^-)^2 \ =\
{\rm
const.}\,\L^{2n^2}
\delta_+\,\delta_-\ ,\ \ {\rm where}\cr \delta_\pm(u_k,\Lambda)\ &=\
\Delta_0(u_2,...,u_{n-1},u_n\pm\L^n)\ ,}
\eqn\DLdef
$$
is the shifted classical discriminant \cdiscdefu. Thus, $\bifset_\L$
consists of two copies of the classical singular locus $\Sigma_0$,
shifted by $\pm\L^n$ in the $u_n$ direction. Obviously, for $\L\to0$,
the classical moduli space is recovered: $\bifset_\L\to \bifset_0$.
That is, when the quantum corrections are switched on, a single
isolated branch of $\bifset_0$ (associated with massless gauge bosons
of a particular $SU(2)$ subgroup) splits into two branches of
$\bifset_\L$ (reflecting two massless Seiberg-Witten dyons related to
this $SU(2)$). This is depicted in \lfig\figsplit.

\figinsert\figsplit{When switching to the exact quantum theory, the
classical singular locus splits into two quantum singularities that
are associated with massless dyons. The distance is governed by the
quantum scale $\L$. }{1.8in}{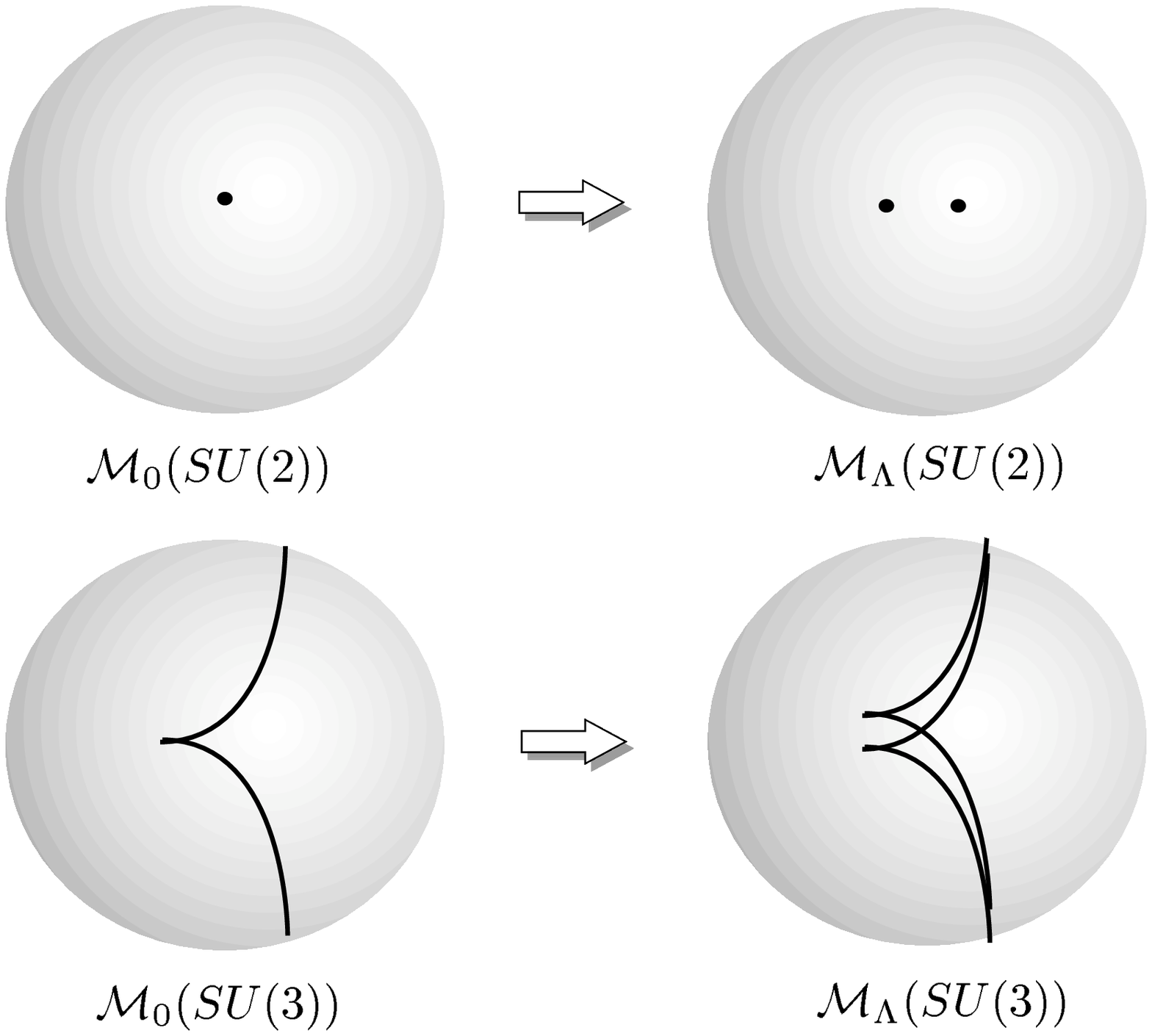}

\ni Moreover, the points $Z_{\l_i}$ of the
classical level surface \levsurf\ split as follows,
$$
Z_{\l_i}(u)\ \to\ Z_{\l_i}^\pm(u,\L) \equiv\
Z_{\l_i}(u_2,,...,u_{n-1},u_n\pm\L^n)\ ,
\eqn\eipm
$$
and become the $2n$ branch points of the Riemann surface \aaa. The
curve can thus be represented by the two-sheeted $x$-plane with cuts
running between pairs $Z_{\l_i}^+$ and $Z_{\l_i}^-$.
See \lfig\figxplane\ for an example.

\figinsert\figxplane{The level manifold of quantum $SU(3)$ Yang-Mills
theory is given by a genus two Riemann surface, which is represented
here as a two-sheeted cover of the $x$-plane. It may be thought as
the quantum version of the classical, zero dimensional level surface
of \lfig\figclassYM, whose points transmute into pairs of branch
points. The dashed lines represent the vanishing cycles (on the upper
sheet) that are associated with the six branches of the singular
locus $\Sigma_\L$. The quantum numbers refer to $(\vec g;\vec q)$,
where $\vec g,\vec q$ are weight vectors in Dynkin basis.
}{1.6in}{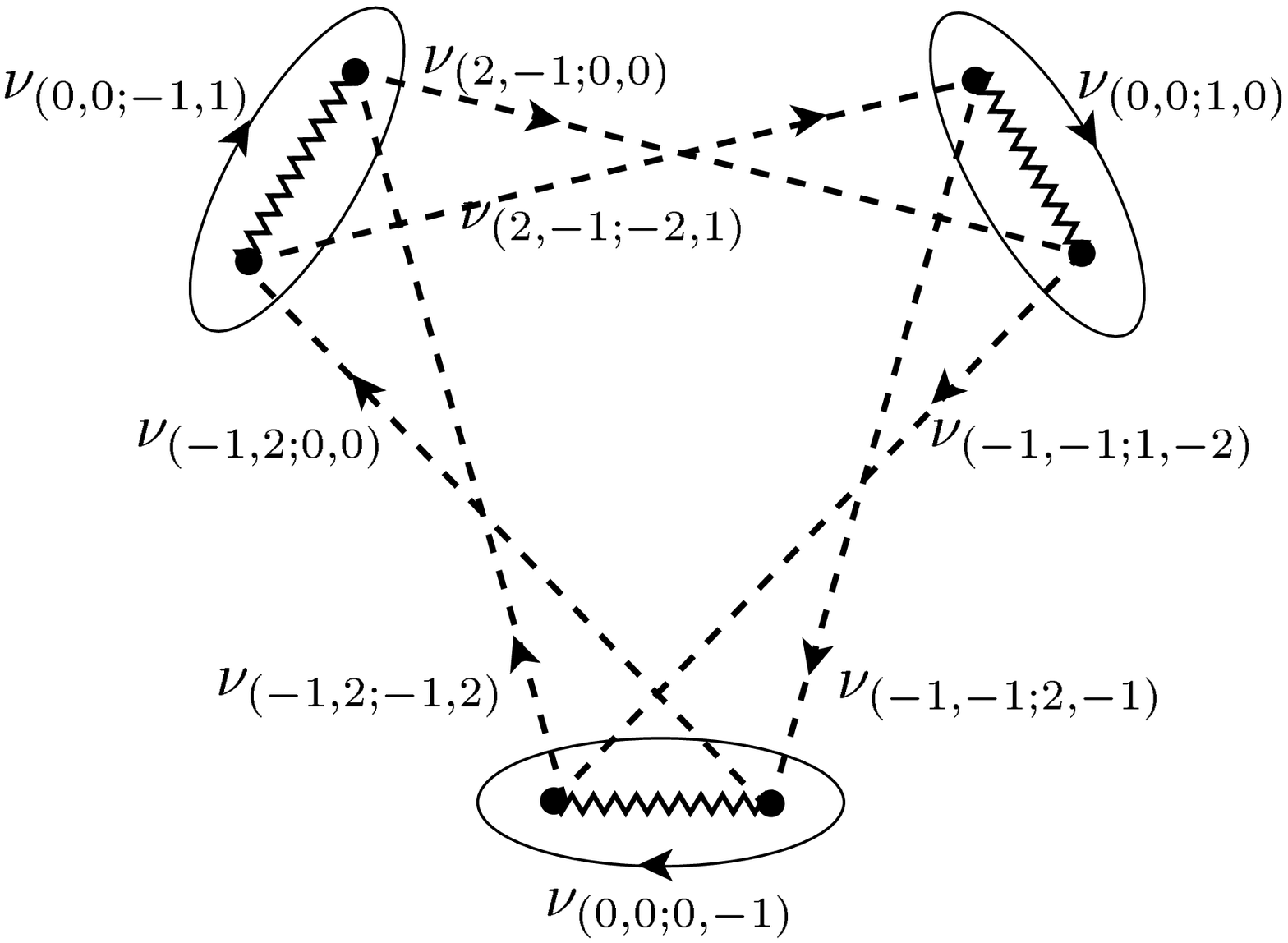}

Just like for the classical level surfaces, the vanishing cycles of
the Riemann surfaces \aaa\ have a concrete group theoretical meaning.
Not only may one expect to determine the quantum numbers of the
massless dyons by just expanding the vanishing cycles in some
appropriate symplectic basis, one find that one can also directly
associate the cycles in the branched $x$-plane with projections of
roots and weights.

Specifically, \lfig\figxplane\ can be thought of as a quantum
deformation
of the classical level surface in \lfig\figclassYM, whose points,
associated with projected weight vectors $\l_i$, turn into branch
cuts (whose length is governed by the quantum scale, $\L$). In fact,
one obtains two, slightly rotated copies of the weight diagram. A
basis of cycles can be chosen such that the coordinates of the
``electric'', $\g_\a$-type of cycles are given by precisely the
weight vectors $\l_i$. Moreover, the classical cycles of
\lfig\figclassYM\ turn into pairs of ``magnetic'', $\g_\b$-type of
cycles, and we can immediately read off the electric and magnetic
quantum numbers of the massless dyons (note that they are given by
root vectors). Accordingly, the intersection properties of the cycles
are reflected by symplectic inner products in the two copies of
weight space.

These considerations apply to $D_n$ gauge groups as well; here, the
relevant quantum surfaces\BL\ are given by analogous deformations of
the classical level surfaces  (cf., \lfig\DsixPolytope),
and one needs to project on $Z_2$-odd cycles.

\goodbreak\vskip1.cm\centerline{{\bf Acknowledgements}}

I would like to thank the organizers of the conference for their hard
work, and A.\ Klemm, S.\ Theisen and S.\ Yankielowicz for enjoyable
collaboration. I also thank A.\ Brandhuber and K.\ Landsteiner for
sharing their insights.
\goodbreak\vskip1.cm
\refout
\end